\documentclass{article}

\PassOptionsToPackage{numbers, compress}{natbib}

     \usepackage[final]{neurips_2024_ml4ps}

\usepackage[utf8]{inputenc} 
\usepackage[T1]{fontenc}    
\usepackage{hyperref}       
\usepackage{url}            
\usepackage{booktabs}       
\usepackage{amsfonts}       
\usepackage{nicefrac}       
\usepackage{microtype}      
\usepackage{xcolor}         

\usepackage[pdftex]{graphicx}
\usepackage{caption}
\usepackage{subcaption}

\title{Cosmological super-resolution of the 21-cm signal}

\author{%
  Simon~Pochinda \\ 
  Cavendish Laboratory\\
  University of Cambridge\\
  J. J. Thomson Avenue, Cambridge, \\CB3 0HE, UK \\
  \texttt{sp2053@cam.ac.uk} \\
   \And
   Jiten Dhandha \\
   Institute of Astronomy\\
   University of Cambridge \\
   Madingley Road, Cambridge, \\CB3 0HA, UK\\
   \AND
  Anastasia Fialkov \\
  Institute of Astronomy\\
  University of Cambridge \\
  Madingley Road, Cambridge, \\CB3 0HA, UK\\
  \And
  Eloy de Lera Acedo \\
  Cavendish Laboratory\\
  University of Cambridge \\
  J. J. Thomson Avenue, Cambridge, \\CB3 0HE, UK \\
}

\begin{document}

\maketitle

\begin{abstract}
In this study, we train score-based diffusion models to super-resolve gigaparsec-scale cosmological simulations of the 21-cm signal. We examine the impact of network and training dataset size on model performance, demonstrating that a single simulation is sufficient for a model to learn the super-resolution task regardless of the initial conditions. Our best-performing model achieves pixelwise \mbox{$\mathrm{RMSE}\sim0.57\ \mathrm{mK}$} and dimensionless power spectrum residuals ranging from $10^{-2}-10^{-1}\ \mathrm{mK^2}$ for $128^3$, $256^3$ and $512^3$ voxel simulation volumes at redshift~$10$. The super-resolution network ultimately allows us to utilize all spatial scales covered by the SKA1-Low instrument, and could in future be employed to help constrain the astrophysics of the early Universe.
\end{abstract}

\section{Introduction}
The Square Kilometre Array \citep[SKA]{dewdney09_ska, schaife20_ska} promises tomographic maps of the 21-cm line of neutral hydrogen along with detailed measurements of summary statistics \citep{furlanetto06_review, pritchard12_review, barkana16_review, mesinger19_review}. The SKA will consist of two phases SKA1 and SKA2 each covering low ($50-350\ \mathrm{MHz}$, $z \approx 3-27$) and mid ($350-15400\ \mathrm{MHz}$) frequencies. Here, we focus on the spatial scales probed by SKA1-Low, which will have frequency-dependent primary station beams covering a field-of-view of approximately $\sim 1\ \mathrm{cGpc}$ at $100\ \mathrm{MHz}$ for $z\sim10$ \citep[Sec. 5.3]{Koopmans15_ska}. Summary statistics of these large cosmological volumes, such as the 21-cm dimensionless power spectrum, will provide valuable information that can be used to constrain astrophysical models of the early Universe \citep[e.g.][]{hera23_idr3,gessey-jones24_cosmicstrings,bevins24_joint,pochinda24_joint}. However, to fully utilize the information contained in the gigaparsec-scales covered by the SKA1-Low, it is necessary to simulate correspondingly large cosmic volumes at high resolution \citep{kaur20_boxsize,ohara24_ska}. This is complicated as producing astrophysical parameter constraints typically requires thousands of simulations to train reliable emulators for inference \citep{bevins24_joint, pochinda24_joint} - which is not feasible, even with state-of-the-art semi-numerical codes.

The emergence of transformer-based architectures \citep{vaswani17_transformers} has improved Convolutional Neural Networks (CNNs) substantially, where diffusion models \citep{sohl15_ddpm,ho20_ddpm, song2021_sm} now beat out Generative Adversarial Networks \citep[GANs,][]{goodfellow14_GANs} in image synthesis \citep{Dhariwal21_beatGANs}. Diffusion models have been employed for a range of tasks including text-to-image generation \citep[e.g. Stable Diffusion,][]{Rombach22_StableDiffusion, podell24_StableDiffusion}, super-resolution \citep[SR,][]{Saharia23_SR3, li22_superres}, image compression \citep{yang23_compression} among others. 
Diffusion models have also been adapted for various applications in astronomy such as galaxy image simulation \citep{smith22_galaxydiffusion}, gravitaitonal lens reconstruction \citep{karchev22_lensrecon}, sampling the initial conditions of the Universe \citep{legin2024_ic}, super-resolving cosmological density fields \citep{Rouhiainen24_inpaintSR}, and even conditional generation of 2D 21-cm differential brightness temperature images \citep{Zhao23_21cmdiffusion}. Based on their superior performance in image generation tasks and successful application in the aforementioned cosmological problems, we likewise adopt diffusion models in this study.  

In this proof of concept work, we train score-based conditional diffusion models \citep{song2021_sm} for super-resolving large 3D cosmological high-resolution (HR) simulations of the 21-cm differential brightness temperature. The SR emulator is aided by conditional inputs consisting of a low-resolution (LR) 21-cm brightness temperature box, and simulation initial conditions. The generative model will help us overcome the computational constraints associated with large-scale HR simulations, which will allow us cover the spatial scales ($k$-modes) probed by the SKA1-Low. 
\section{Methods}
\subsection{Simulations of the 21-cm line}\label{sec:simulations}
The simulations used in this SR study are semi-numerical simulations of the 21-cm differential brightness temperature generated by the code \textsc{21cmSPACE} \citep{Visbal_2012, Fialkov_2012, Fialkov_2013, Fialkov_2014, Fialkov_2014b, Cohen_2016, Fialkov_2019, Reis_2020, Reis_2021, Reis_2022, Magg_2022, Gessey-Jones_2022, Gessey-Jones_2023, Sikder_2023}. The dataset used in this work consists of 80 HR simulations of $768^3\ \mathrm{cMpc}^3$ volumes with $256^3$ voxels each $3\ \mathrm{cMpc}$ across. While the simulations cover integer redshifts from $z=6-49$, we focus on $z=10$ and leave redshift conditional generation to a future study. The simulations are all run with the same astrophysical and cosmological parameters, but different initial conditions. These initial conditions consists of matter overdensity fields and baryon-dark matter relative velocity fields also spanning $768^3\ \mathrm{cMpc}^3$ with $256^3$ voxels at $3\ \mathrm{cMpc}$ resolution. The conditional LR input is generated from the HR simulations by a trilinear downsampling to $12\ \mathrm{cMpc}$ voxels, followed by a trilinear upsampling to match the other input dimensions. The 80 simulations in the dataset took $\sim 34000$ CPU hours to complete ($20$ CPUs for $\sim20$ hours for each simulation).

Additionally, we run a single large simulation to evaluate the model on the large spatial scales covered by SKA1-Low. This large simulation covers a $1536^3\ \mathrm{cMpc}^3$ volume with $512^3$ voxels at $3\ \mathrm{cMpc}$ resolution from $z=6-49$. In our analysis of the performance across spatial scales, we likewise choose to focus on $z=10$. The large simulation took $\sim45000$ CPU hours (128 CPUs for $\sim2$ weeks runtime) to run.

\subsection{Super-resolution network architecture and training procedure}
We train score-based Variance Preserving Stochastic Differential Equation (VP SDE) diffusion models using the noise-weighted training objective as described in \citet{song2021_sm}. 
We utilize the improved \textsc{DDPM++} architecture for the denoising network, as implemented by \citet{karras22_elucidating}, adapting it to handle 3D inputs. The models in this work use 4 residual blocks per resolution, resolution multipliers 1, 2, 4, 8, and 16 for the number of channels, and self-attention at the $8^3$ and bottleneck resolution \citep{vaswani17_transformers,Wang18_selfattention}. 
For sampling, we employ the Euler-Maruyama algorithm with 100 steps for the reverse diffusion process. While we use the stochastic Euler-Maruyama sampler, the simulated data in this work does not include any inherent stochasticity. Thus, deterministic sampling methods like ODE-based solvers could also be explored for inference. However, 21cmSPACE can model stochastic processes which might benefit from the noise injected at every reverse diffusion step. Hence our choice of stochastic sampling is compatible with potential future extensions including stochastic processes in the simulations.

The training procedure for an epoch can be broken down into the following: A $256^3$ HR simulation is drawn from the training dataset, along with the corresponding initial conditions. To accelerate training, each HR cube is split into $8\times 128^3$ smaller cubes to train on two batches of 4 simulations. The conditional cubes are normalized by subtracting their mean and dividing by their standard deviation. As the statistics of the HR cube are unknown in a realistic scenario, the HR target is normalized to the mean and standard deviation of the downsampled LR cube. Input cubes are augmented by applying one of 24 unique rotations, increasing the effective size of our training dataset. The HR cube is attenuated according to the drawn noise level. The initial conditions and LR input are then concatenated to the noisy HR cube. A forward pass through the model is performed, followed by backpropagation to update the model weights.

\begin{figure}
\centering
\includegraphics[width=\linewidth]{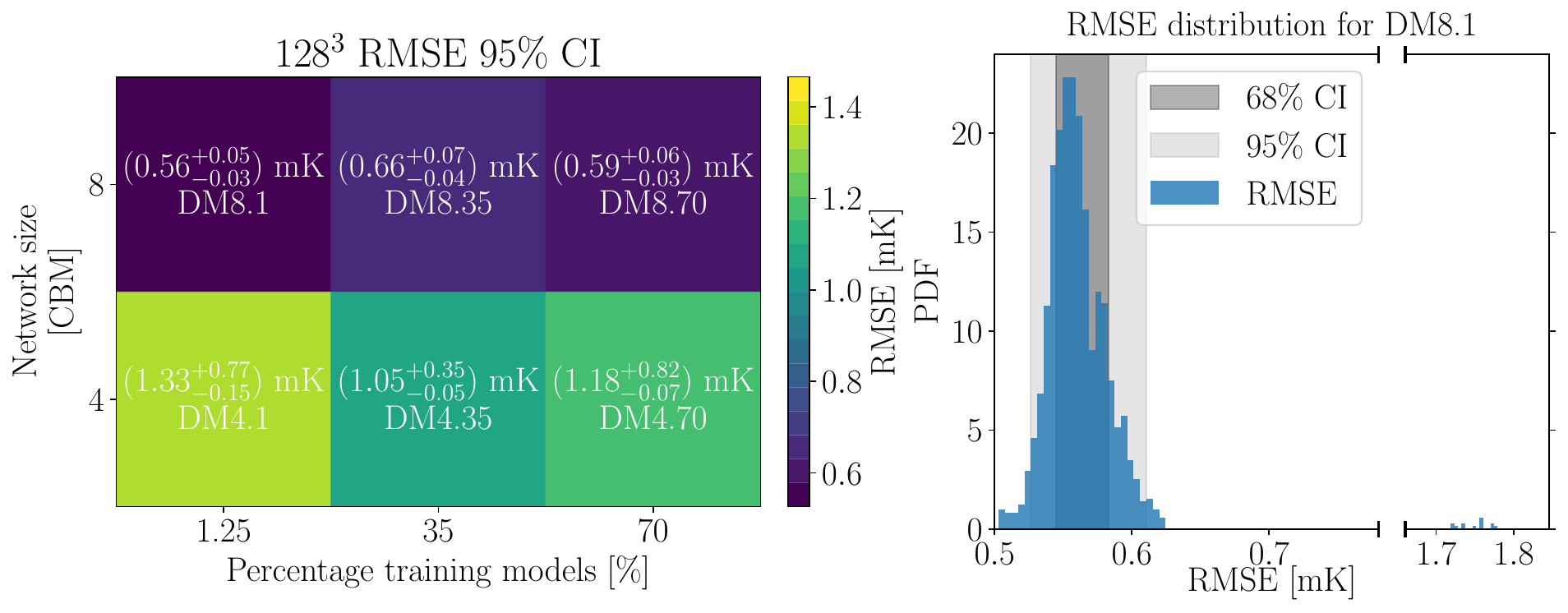}
\caption{\textit{Left}: Models of varying network size ($\mathrm{CBM}\in \{4,8\}$) trained on 1.25\%/35\%/70\% (1/28/56 models) of the dataset. \textit{Right}: RMSE distribution of the model with $\mathrm{CBM}=8$ and trained on 1.25\% of the dataset.}\label{fig:vary_rmse}
\end{figure}

We train models using the Adam optimizer \citep{kingma14_adam} with an initial learning rate of $10^{-3}$. After the first 1000 epochs, the learning rate is reduced to $10^{-4}$. Validation checks are conducted every 50 epochs after the model passes 1000 epochs of training. During validation, a $256^3$ simulation and initial conditions are drawn from the validation set, and a sample from the model is generated. A checkpoint is saved if the pixelwise Root Mean Squared Error (RMSE) of the validation sample reaches a new minimum. The model is trained for $\sim12$ hours on four NVIDIA A100-SXM-80GB GPUs.

\section{Results}\label{sec:results}
\subsection{Effect of Network and Training Dataset Size}
The dataset is split into three different training, validation, and test percentage configurations i.e.: (training, validation, test)\% = $(70, 20, 10)$\%, $(35, 55, 10)$\%, and $(1.25, 88.75, 10)$\%. In terms of number of simulations for training, validation and testing, this corresponds to $(56, 16, 8)$ simulations, $(28, 44, 8)$ simulations, and \mbox{$(1, 71, 8)$} simulations respectively. Additionally, we train two network configurations -- a small network with a channel base multiplier (CBM) of 4, and a large network with a $\mathrm{CBM}=8$. With the two different network sizes and the three different dataset splits, a total of 6 models are trained. This setup allows us to examine how the size of the training dataset and the network architecture impact model performance. Each model is referred to according to the following convention: e.g. DM8.1 is the diffusion model (DM) with $\mathrm{CBM}=8$ and a rounded $1.25\%$ data for training. For the post-training analysis we split the 10\% (8 models) test data into $64\times 128^3$ cubes, and rotate each simulation 24 unique times to effectively evaluate the network on $1536\times 128^3$ boxes. 

The left-hand side of Figure~\ref{fig:vary_rmse} shows the 95 percentile confidence interval (CI) for the pixelwise-RMSE across combinations of network size (CBM) and training dataset size. While the pixelwise-RMSE decreases with network size (from $\sim1.2$ mK to $\sim0.6$ mK), the overall noise level remains far below the error margins of current 21-cm experiments specialized in measuring the global 21-cm signal, such as REACH \citep[$25\ \mathrm{mK}$\ RMSE]{Eloy22_reach} and SARAS~3 \citep[$213 \mathrm{mK}$ RMSE]{Singh22_saras3}. Models with $\mathrm{CBM}=4$ showed higher variance in the pixelwise-RMSE with the training dataset size, however all 95\% CIs overlap significantly, showing no strong dependency on the size of the training dataset. For the larger networks ($\mathrm{CBM}=8$), the pixelwise-RMSE variance is smaller and more stable across training dataset size, suggesting the network performance is independent of the amount of training data. Hence, training a SR network for a set of astrophysical parameters is computationally feasible as only a single simulation is required to learn the high-redshift astrophysics regardless of the initial conditions. The right-hand side of Figure~\ref{fig:vary_rmse} shows the pixelwise-RMSE distribution for the $1536$ test models sampled by DM8.1 along with 68\% and 95\% CIs. The majority of samples ($99.15\%$) have \mbox{$\mathrm{RMSE}<0.63\ \mathrm{mK}$} while just a few outliers ($0.85\%$) have \mbox{$\mathrm{RMSE}\approx1.75\ \mathrm{mK}$}. 

While learning the super-resolution task from a single training simulation is a significant result, it is worth noting that each simulation contains millions of voxels. The physical size of the training simulations is large enough to neglect cosmic variance, which means the summary statistics of the simulations are entirely determined by the astrophysical and cosmological parameters. Since these parameters are identical for all simulations, this may explain why a single simulation is sufficient to train a super-resolution network.

\begin{figure}
\centering
\includegraphics[width=0.8\linewidth]{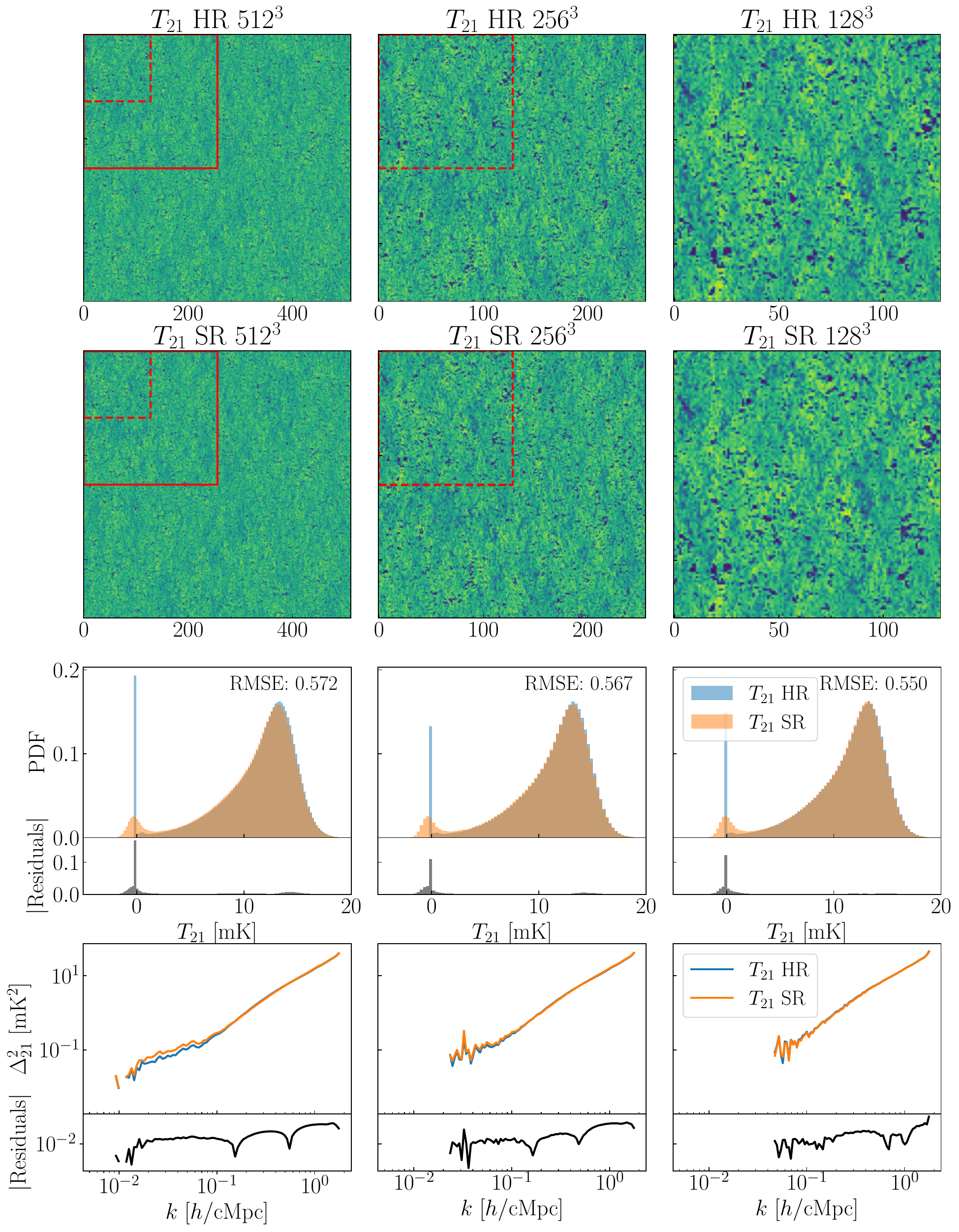}
\caption{Row 1 shows slices from the 21-cm HR $512^3$, $256^3$, and $128^3$ targets. Row 2 shows slices from sampled models at each scale using DM8.1. Each sample is independently generated from the model, rather than being zoomed-in sections of the largest cube. Row 3 presents histograms of the voxel distributions, and Row 4 compares the 21-cm dimensionless power spectrum of HR targets and SR predictions across different spatial scales.}\label{fig:scales}
\end{figure}

\subsection{Super-resolving cosmological volumes at different scales}
Following the evaluation of model and training dataset sizes, we tested the best-performing model, DM8.1, across different spatial scales. For this analysis we used a large $1536^3\ \mathrm{cMpc}^3$ simulation as described in Section~\ref{sec:simulations}. Figure~\ref{fig:scales} shows simulation and model slices and statistics for $512^3$, $256^3$, and $128^3$ spatial scales. Specifically, the rows show target 21-cm HR simulation slices, sampled SR model slices, histograms of pixel probability distribution functions (PDF), and 21-cm dimensionless power spectra. It is worth emphasizing that the SR output in each column of Figure~\ref{fig:scales} is independently sampled from the model (i.e. the SR row is not simply crop-outs of the larger box). 
On the largest scale DM8.1 achieved an \mbox{$\mathrm{RMSE_{512^3}}=0.57_{-0.02}^{+0.02}\ \mathrm{mK}$} on 24 rotations of the simulated box. To test intermediate scales, we further divided the large simulation into $8\times 256^3$ cubes, on which DM8.1 achieved an $\mathrm{RMSE_{256^3}} = 0.56_{-0.01}^{+0.03}\ \mathrm{mK}$. Finally, on the smallest scale of $128^3$ (also previously shown in Figure~\ref{fig:vary_rmse}), DM8.1 achieved an $\mathrm{RMSE_{128^3}}=0.56^{+0.05}_{-0.03}\ \mathrm{mK}$. DM8.1 also demonstrated residuals within the $10^{-2}-10^{-1}\ \mathrm{mK^2}$ range for the dimensionless power spectrum, $\Delta^2_\mathrm{21}$, which is below the expected noise level of SKA1-Low \citep{kaur20_boxsize}. 

The pixel histograms in Figure~\ref{fig:scales} also reveal worse performance around the sharp peak at $T_{21}=0\ \mathrm{mK}$, which corresponds to fully ionized regions where the 21-cm line is extinguished.   
In our training simulations the 21-cm signal is only observed in emission (i.e. $T_{21}\geq0$). Consequently, one could explore adding a positivity constraint by learning a quantitiy where e.g. $\exp(T_{21}^\ast)=T_{21}$, which might reduce some of the observed discrepancies in the PDF in Figure~\ref{fig:scales}. However, in reality and in the simulations, the 21-cm signal can assume both negative and positive values. The sign of the signal depends on the interplay between the initial conditions and the astrophysical processes as well as the redshift and spatial scales observed. As we ultimately aim to develop a methodology that generalizes across a range of redshifts, including those where $T_{21}$ can also assume negative values, we do not adopt any positivity constraints as that would limit the applicability of our approach. 

The SR network offers a significant reduction in computational resources taking just 4.7 hours to sample all 24 unique rotations of the $512^3$ simulation cube on four NVIDIA A100-SXM-80GB GPUs.

\section{Conclusions and Future Work}
In this work, we trained score-based diffusion models to super-resolve gigaparsec-scale cosmological simulations of the 21-cm differential brightness temperature. 
We explored the impact of network architecture and training dataset size on the performance of trained SR networks. We demonstrated that the model performance in terms of the pixelwise-RMSE is independent of the training dataset size, i.e. a single simulation is enough to learn the data distribution regardless of the initial conditions. This is likely due to our training simulation volumes being large enough to neglect cosmic variance, such that the statistics are solely determined by the astrophysical and cosmological parameters, which are identical for all simulations. 
Moreover, we present sub-mK performance for the large models, which is well within the error margins of current 21-cm experiments \citep{Eloy22_reach, Singh22_saras3}.

Finally, we examined our best-performing model across a range of spatial scales. We found \mbox{$\mathrm{RMSE}\sim0.57\ \mathrm{mK}$} across sampled $512^3$, $256^3$, and $128^3$ simulation volumes, and in terms of the dimensionless power spectrum the residuals were of the order $10^{-2}-10^{-1}\ \mathrm{mK^2}$, which is below the expected SKA1-Low noise level \citep{kaur20_boxsize}. Additionally, our trained models successfully match the full probability distribution function (PDF) while achieving a low pixelwise-RMSE and low power spectrum residuals. While we believe jointly demonstrating a low RMSE, matching the power spectrum and the PDF, constitutes a challenging task in itself, and underscores the capabilities of the trained models, the model performance could also be evaluated based on higher-order statistics such as the bispectrum, which we will explore in future work. 

In terms of future extensions to this study, training models with varying amounts of conditional information could also be explored. This could be attempted by systematically removing either of the initial conditions or the low-resolution input. In this proof-of-concept study, we provide the model with both the initial conditions and a low-resolution input. For this initial investigation, we considered the LR inputs essential, as they contain unique information about the astrophysical processes at play. These processes are independent from the initial conditions and are highly uncertain. Therefore, it would be extremely challenging - if not impossible - to reconstruct the 21-cm signal from the initial conditions without providing this extra information.

Finally, future extensions to speed up sampling, and add parameter-conditional generation should be explored. Faster sampling could be attempted using a methodology similar to \citet{karras22_elucidating}, and parameter-conditional generation could be implemented using a latent diffusion model architecture \citep{vahdat21_sbldm, Rombach22_StableDiffusion}.
Parameter-conditioned super-resolution would allow us to use the largest spatial scales probed by the SKA1-Low to constrain astrophysical models, and infer the properties of the early Universe. 

\section*{Acknowledgements}
SP acknowledges the support from the Cambridge Commonwealth, European \& International Trust in the form of a scholarship to undertake his degree and support this paper. JD acknowledges support from the Boustany Foundation and Cambridge Commonwealth Trust in the form of an Isaac Newton Studentship. AF is grateful for the support from the Royal Society through a University Research Fellowship. EdLA is supported by the Science and Technology Facilities Council (UK) through his Rutherford Fellowship.

This work used the DiRAC COSMA Durham facility managed by the Institute for Computational Cosmology on behalf of the STFC DiRAC HPC Facility (www.dirac.ac.uk). The equipment was funded by BEIS capital funding via STFC capital grants ST/P002293/1, ST/R002371/1 and ST/S002502/1, Durham University and STFC operations grant ST/R000832/1. DiRAC is part of the National e-Infrastructure.

This work was also supported by a seedcorn project which used the DiRAC Extreme Scaling service Tursa at the University of Edinburgh, managed by the Edinburgh Parallel Computing Centre on behalf of the STFC DiRAC HPC Facility (www.dirac.ac.uk). The DiRAC service at Edinburgh was funded by BEIS, UKRI and STFC capital funding and STFC operations grants. DiRAC is part of the UKRI Digital Research Infrastructure

Finally, this work was also performed using resources provided by the Cambridge Service for Data Driven Discovery (CSD3) operated by the University of Cambridge Research Computing Service (www.csd3.cam.ac.uk), provided by Dell EMC and Intel using Tier-2 funding from the Engineering and Physical Sciences Research Council (capital grant EP/T022159/1), and DiRAC funding from the Science and Technology Facilities Council (www.dirac.ac.uk).
\bibliographystyle{abbrvnat}
\bibliography{main}

\begin{thebibliography}{49}
\providecommand{\natexlab}[1]{#1}
\providecommand{\url}[1]{\texttt{#1}}
\expandafter\ifx\csname urlstyle\endcsname\relax
  \providecommand{\doi}[1]{doi: #1}\else
  \providecommand{\doi}{doi: \begingroup \urlstyle{rm}\Url}\fi

\bibitem[{Barkana}(2016)]{barkana16_review}
R.~{Barkana}.
\newblock {The rise of the first stars: Supersonic streaming, radiative feedback, and 21-cm cosmology}.
\newblock \emph{\physrep}, 645:\penalty0 1--59, July 2016.
\newblock \doi{10.1016/j.physrep.2016.06.006}.

\bibitem[{Bevins} et~al.(2024){Bevins}, {Heimersheim}, {Abril-Cabezas}, {Fialkov}, {de Lera Acedo}, {Handley}, {Singh}, and {Barkana}]{bevins24_joint}
H.~T.~J. {Bevins}, S.~{Heimersheim}, I.~{Abril-Cabezas}, A.~{Fialkov}, E.~{de Lera Acedo}, W.~{Handley}, S.~{Singh}, and R.~{Barkana}.
\newblock {Joint analysis constraints on the physics of the first galaxies with low-frequency radio astronomy data}.
\newblock \emph{\mnras}, 527\penalty0 (1):\penalty0 813--827, Jan. 2024.
\newblock \doi{10.1093/mnras/stad3194}.

\bibitem[{Cohen} et~al.(2016){Cohen}, {Fialkov}, and {Barkana}]{Cohen_2016}
A.~{Cohen}, A.~{Fialkov}, and R.~{Barkana}.
\newblock {The 21-cm BAO signature of enriched low-mass galaxies during cosmic reionization}.
\newblock \emph{\mnras}, 459\penalty0 (1):\penalty0 L90--L94, June 2016.
\newblock \doi{10.1093/mnrasl/slw047}.

\bibitem[{de Lera Acedo} et~al.(2022){de Lera Acedo}, {de Villiers}, {Razavi-Ghods}, {Handley}, {Fialkov}, {Magro}, {Anstey}, {Bevins}, {Chiello}, {Cumner}, {Josaitis}, {Roque}, {Sims}, {Scheutwinkel}, {Alexander}, {Bernardi}, {Carey}, {Cavillot}, {Croukamp}, {Ely}, {Gessey-Jones}, {Gueuning}, {Hills}, {Kulkarni}, {Maiolino}, {Meerburg}, {Mittal}, {Pritchard}, {Puchwein}, {Saxena}, {Shen}, {Smirnov}, {Spinelli}, and {Zarb-Adami}]{Eloy22_reach}
E.~{de Lera Acedo}, D.~I.~L. {de Villiers}, N.~{Razavi-Ghods}, W.~{Handley}, A.~{Fialkov}, A.~{Magro}, D.~{Anstey}, H.~T.~J. {Bevins}, R.~{Chiello}, J.~{Cumner}, A.~T. {Josaitis}, I.~L.~V. {Roque}, P.~H. {Sims}, K.~H. {Scheutwinkel}, P.~{Alexander}, G.~{Bernardi}, S.~{Carey}, J.~{Cavillot}, W.~{Croukamp}, J.~A. {Ely}, T.~{Gessey-Jones}, Q.~{Gueuning}, R.~{Hills}, G.~{Kulkarni}, R.~{Maiolino}, P.~D. {Meerburg}, S.~{Mittal}, J.~R. {Pritchard}, E.~{Puchwein}, A.~{Saxena}, E.~{Shen}, O.~{Smirnov}, M.~{Spinelli}, and K.~{Zarb-Adami}.
\newblock {The REACH radiometer for detecting the 21-cm hydrogen signal from redshift z {\ensuremath{\approx}} 7.5-28}.
\newblock \emph{Nature Astronomy}, 6:\penalty0 984--998, July 2022.
\newblock \doi{10.1038/s41550-022-01709-9}.

\bibitem[Dewdney et~al.(2009)Dewdney, Hall, Schilizzi, and Lazio]{dewdney09_ska}
P.~E. Dewdney, P.~J. Hall, R.~T. Schilizzi, and T.~J. L.~W. Lazio.
\newblock The square kilometre array.
\newblock \emph{Proceedings of the IEEE}, 97\penalty0 (8):\penalty0 1482--1496, 2009.
\newblock \doi{10.1109/JPROC.2009.2021005}.

\bibitem[Dhariwal and Nichol(2021)]{Dhariwal21_beatGANs}
P.~Dhariwal and A.~Nichol.
\newblock Diffusion models beat gans on image synthesis.
\newblock In M.~Ranzato, A.~Beygelzimer, Y.~Dauphin, P.~Liang, and J.~W. Vaughan, editors, \emph{Advances in Neural Information Processing Systems}, volume~34 of \emph{NIPS 2021}, pages 8780--8794. Curran Associates, Inc., 2021.
\newblock URL \url{https://proceedings.neurips.cc/paper_files/paper/2021/file/49ad23d1ec9fa4bd8d77d02681df5cfa-Paper.pdf}.

\bibitem[{Fialkov} and {Barkana}(2019)]{Fialkov_2019}
A.~{Fialkov} and R.~{Barkana}.
\newblock {Signature of excess radio background in the 21-cm global signal and power spectrum}.
\newblock \emph{\mnras}, 486\penalty0 (2):\penalty0 1763--1773, June 2019.
\newblock \doi{10.1093/mnras/stz873}.

\bibitem[{Fialkov} et~al.(2012){Fialkov}, {Barkana}, {Tseliakhovich}, and {Hirata}]{Fialkov_2012}
A.~{Fialkov}, R.~{Barkana}, D.~{Tseliakhovich}, and C.~M. {Hirata}.
\newblock {Impact of the relative motion between the dark matter and baryons on the first stars: semi-analytical modelling}.
\newblock \emph{\mnras}, 424\penalty0 (2):\penalty0 1335--1345, Aug. 2012.
\newblock \doi{10.1111/j.1365-2966.2012.21318.x}.

\bibitem[{Fialkov} et~al.(2013){Fialkov}, {Barkana}, {Visbal}, {Tseliakhovich}, and {Hirata}]{Fialkov_2013}
A.~{Fialkov}, R.~{Barkana}, E.~{Visbal}, D.~{Tseliakhovich}, and C.~M. {Hirata}.
\newblock {The 21-cm signature of the first stars during the Lyman-Werner feedback era}.
\newblock \emph{\mnras}, 432\penalty0 (4):\penalty0 2909--2916, July 2013.
\newblock \doi{10.1093/mnras/stt650}.

\bibitem[{Fialkov} et~al.(2014{\natexlab{a}}){Fialkov}, {Barkana}, {Pinhas}, and {Visbal}]{Fialkov_2014}
A.~{Fialkov}, R.~{Barkana}, A.~{Pinhas}, and E.~{Visbal}.
\newblock {Complete history of the observable 21 cm signal from the first stars during the pre-reionization era}.
\newblock \emph{\mnras}, 437\penalty0 (1):\penalty0 L36--L40, Jan. 2014{\natexlab{a}}.
\newblock \doi{10.1093/mnrasl/slt135}.

\bibitem[{Fialkov} et~al.(2014{\natexlab{b}}){Fialkov}, {Barkana}, and {Visbal}]{Fialkov_2014b}
A.~{Fialkov}, R.~{Barkana}, and E.~{Visbal}.
\newblock {The observable signature of late heating of the Universe during cosmic reionization}.
\newblock \emph{\nat}, 506\penalty0 (7487):\penalty0 197--199, Feb. 2014{\natexlab{b}}.
\newblock \doi{10.1038/nature12999}.

\bibitem[{Furlanetto} et~al.(2006){Furlanetto}, {Oh}, and {Briggs}]{furlanetto06_review}
S.~R. {Furlanetto}, S.~P. {Oh}, and F.~H. {Briggs}.
\newblock {Cosmology at low frequencies: The 21 cm transition and the high-redshift Universe}.
\newblock \emph{\physrep}, 433\penalty0 (4-6):\penalty0 181--301, Oct. 2006.
\newblock \doi{10.1016/j.physrep.2006.08.002}.

\bibitem[{Gessey-Jones} et~al.(2022){Gessey-Jones}, {Sartorio}, {Fialkov}, {Mirouh}, {Magg}, {Izzard}, {de Lera Acedo}, {Handley}, and {Barkana}]{Gessey-Jones_2022}
T.~{Gessey-Jones}, N.~S. {Sartorio}, A.~{Fialkov}, G.~M. {Mirouh}, M.~{Magg}, R.~G. {Izzard}, E.~{de Lera Acedo}, W.~J. {Handley}, and R.~{Barkana}.
\newblock {Impact of the primordial stellar initial mass function on the 21-cm signal}.
\newblock \emph{\mnras}, 516\penalty0 (1):\penalty0 841--860, Oct. 2022.
\newblock \doi{10.1093/mnras/stac2049}.

\bibitem[{Gessey-Jones} et~al.(2023){Gessey-Jones}, {Fialkov}, {de Lera Acedo}, {Handley}, and {Barkana}]{Gessey-Jones_2023}
T.~{Gessey-Jones}, A.~{Fialkov}, E.~{de Lera Acedo}, W.~J. {Handley}, and R.~{Barkana}.
\newblock {Signatures of cosmic ray heating in 21-cm observables}.
\newblock \emph{\mnras}, 526\penalty0 (3):\penalty0 4262--4284, Dec. 2023.
\newblock \doi{10.1093/mnras/stad3014}.

\bibitem[{Gessey-Jones} et~al.(2024){Gessey-Jones}, {Pochinda}, {Bevins}, {Fialkov}, {Handley}, {de Lera Acedo}, {Singh}, and {Barkana}]{gessey-jones24_cosmicstrings}
T.~{Gessey-Jones}, S.~{Pochinda}, H.~T.~J. {Bevins}, A.~{Fialkov}, W.~J. {Handley}, E.~{de Lera Acedo}, S.~{Singh}, and R.~{Barkana}.
\newblock {On the constraints on superconducting cosmic strings from 21-cm cosmology}.
\newblock \emph{\mnras}, 529\penalty0 (1):\penalty0 519--536, Mar. 2024.
\newblock \doi{10.1093/mnras/stae512}.

\bibitem[Goodfellow et~al.(2014)Goodfellow, Pouget-Abadie, Mirza, Xu, Warde-Farley, Ozair, Courville, and Bengio]{goodfellow14_GANs}
I.~Goodfellow, J.~Pouget-Abadie, M.~Mirza, B.~Xu, D.~Warde-Farley, S.~Ozair, A.~Courville, and Y.~Bengio.
\newblock Generative adversarial nets.
\newblock In Z.~Ghahramani, M.~Welling, C.~Cortes, N.~Lawrence, and K.~Weinberger, editors, \emph{Advances in Neural Information Processing Systems}, volume~27 of \emph{NIPS 2014}. Curran Associates, Inc., 2014.
\newblock URL \url{https://proceedings.neurips.cc/paper_files/paper/2014/file/5ca3e9b122f61f8f06494c97b1afccf3-Paper.pdf}.

\bibitem[{HERA Collaboration} et~al.(2023){HERA Collaboration}, {Abdurashidova}, {Adams}, {Aguirre}, {Alexander}, {Ali}, {Baartman}, {Balfour}, {Barkana}, {Beardsley}, {Bernardi}, {Billings}, {Bowman}, {Bradley}, {Breitman}, {Bull}, {Burba}, {Carey}, {Carilli}, {Cheng}, {Choudhuri}, {DeBoer}, {de Lera Acedo}, {Dexter}, {Dillon}, {Ely}, {Ewall-Wice}, {Fagnoni}, {Fialkov}, {Fritz}, {Furlanetto}, {Gale-Sides}, {Garsden}, {Glendenning}, {Gorce}, {Gorthi}, {Greig}, {Grobbelaar}, {Halday}, {Hazelton}, {Heimersheim}, {Hewitt}, {Hickish}, {Jacobs}, {Julius}, {Kern}, {Kerrigan}, {Kittiwisit}, {Kohn}, {Kolopanis}, {Lanman}, {La Plante}, {Lewis}, {Liu}, {Loots}, {Ma}, {MacMahon}, {Malan}, {Malgas}, {Malgas}, {Maree}, {Marero}, {Martinot}, {McBride}, {Mesinger}, {Mirocha}, {Molewa}, {Morales}, {Mosiane}, {Mu{\~n}oz}, {Murray}, {Nagpal}, {Neben}, {Nikolic}, {Nunhokee}, {Nuwegeld}, {Parsons}, {Pascua}, {Patra}, {Pieterse}, {Qin}, {Razavi-Ghods}, {Robnett}, {Rosie}, {Santos}, {Sims}, {Singh}, {Smith}, {Swarts}, {Tan},
  {Thyagarajan}, {Wilensky}, {Williams}, {van Wyngaarden}, and {Zheng}]{hera23_idr3}
{HERA Collaboration}, Z.~{Abdurashidova}, T.~{Adams}, J.~E. {Aguirre}, P.~{Alexander}, Z.~S. {Ali}, R.~{Baartman}, Y.~{Balfour}, R.~{Barkana}, A.~P. {Beardsley}, G.~{Bernardi}, T.~S. {Billings}, J.~D. {Bowman}, R.~F. {Bradley}, D.~{Breitman}, P.~{Bull}, J.~{Burba}, S.~{Carey}, C.~L. {Carilli}, C.~{Cheng}, S.~{Choudhuri}, D.~R. {DeBoer}, E.~{de Lera Acedo}, M.~{Dexter}, J.~S. {Dillon}, J.~{Ely}, A.~{Ewall-Wice}, N.~{Fagnoni}, A.~{Fialkov}, R.~{Fritz}, S.~R. {Furlanetto}, K.~{Gale-Sides}, H.~{Garsden}, B.~{Glendenning}, A.~{Gorce}, D.~{Gorthi}, B.~{Greig}, J.~{Grobbelaar}, Z.~{Halday}, B.~J. {Hazelton}, S.~{Heimersheim}, J.~N. {Hewitt}, J.~{Hickish}, D.~C. {Jacobs}, A.~{Julius}, N.~S. {Kern}, J.~{Kerrigan}, P.~{Kittiwisit}, S.~A. {Kohn}, M.~{Kolopanis}, A.~{Lanman}, P.~{La Plante}, D.~{Lewis}, A.~{Liu}, A.~{Loots}, Y.-Z. {Ma}, D.~H.~E. {MacMahon}, L.~{Malan}, K.~{Malgas}, C.~{Malgas}, M.~{Maree}, B.~{Marero}, Z.~E. {Martinot}, L.~{McBride}, A.~{Mesinger}, J.~{Mirocha}, M.~{Molewa}, M.~F. {Morales},
  T.~{Mosiane}, J.~B. {Mu{\~n}oz}, S.~G. {Murray}, V.~{Nagpal}, A.~R. {Neben}, B.~{Nikolic}, C.~D. {Nunhokee}, H.~{Nuwegeld}, A.~R. {Parsons}, R.~{Pascua}, N.~{Patra}, S.~{Pieterse}, Y.~{Qin}, N.~{Razavi-Ghods}, J.~{Robnett}, K.~{Rosie}, M.~G. {Santos}, P.~{Sims}, S.~{Singh}, C.~{Smith}, H.~{Swarts}, J.~{Tan}, N.~{Thyagarajan}, M.~J. {Wilensky}, P.~K.~G. {Williams}, P.~{van Wyngaarden}, and H.~{Zheng}.
\newblock {Improved Constraints on the 21 cm EoR Power Spectrum and the X-Ray Heating of the IGM with HERA Phase I Observations}.
\newblock \emph{\apj}, 945\penalty0 (2):\penalty0 124, Mar. 2023.
\newblock \doi{10.3847/1538-4357/acaf50}.

\bibitem[Ho et~al.(2020)Ho, Jain, and Abbeel]{ho20_ddpm}
J.~Ho, A.~Jain, and P.~Abbeel.
\newblock Denoising diffusion probabilistic models.
\newblock In H.~Larochelle, M.~Ranzato, R.~Hadsell, M.~Balcan, and H.~Lin, editors, \emph{Advances in Neural Information Processing Systems}, volume~33 of \emph{NIPS 2020}, pages 6840--6851. Curran Associates, Inc., 2020.
\newblock URL \url{https://proceedings.neurips.cc/paper_files/paper/2020/file/4c5bcfec8584af0d967f1ab10179ca4b-Paper.pdf}.

\bibitem[Karchev et~al.(2022)Karchev, Anau~Montel, Coogan, and Weniger]{karchev22_lensrecon}
K.~Karchev, N.~Anau~Montel, A.~Coogan, and C.~Weniger.
\newblock {Strong-Lensing Source Reconstruction with Denoising Diffusion Restoration Models}.
\newblock In \emph{{36th Conference on Neural Information Processing Systems}: {Workshop on Machine Learning and the Physical Sciences}}, 11 2022.

\bibitem[Karras et~al.(2022)Karras, Aittala, Aila, and Laine]{karras22_elucidating}
T.~Karras, M.~Aittala, T.~Aila, and S.~Laine.
\newblock Elucidating the design space of diffusion-based generative models.
\newblock In S.~Koyejo, S.~Mohamed, A.~Agarwal, D.~Belgrave, K.~Cho, and A.~Oh, editors, \emph{Advances in Neural Information Processing Systems}, volume~35, pages 26565--26577. Curran Associates, Inc., 2022.
\newblock URL \url{https://proceedings.neurips.cc/paper_files/paper/2022/file/a98846e9d9cc01cfb87eb694d946ce6b-Paper-Conference.pdf}.

\bibitem[Kaur et~al.(2020)Kaur, Gillet, and Mesinger]{kaur20_boxsize}
H.~D. Kaur, N.~Gillet, and A.~Mesinger.
\newblock {Minimum size of 21-cm simulations}.
\newblock \emph{Monthly Notices of the Royal Astronomical Society}, 495\penalty0 (2):\penalty0 2354--2362, 05 2020.
\newblock ISSN 0035-8711.
\newblock \doi{10.1093/mnras/staa1323}.
\newblock URL \url{https://doi.org/10.1093/mnras/staa1323}.

\bibitem[{Kingma} and {Ba}(2014)]{kingma14_adam}
D.~P. {Kingma} and J.~{Ba}.
\newblock {Adam: A Method for Stochastic Optimization}.
\newblock \emph{arXiv e-prints}, art. arXiv:1412.6980, Dec. 2014.
\newblock \doi{10.48550/arXiv.1412.6980}.

\bibitem[Koopmans et~al.(2015)Koopmans, Pritchard, Mellema, Aguirre, Ahn, Barkana, van Bemmel, Bernardi, Bonaldi, Briggs, de~Bruyn, Chang, Chapman, Chen, Courty, Dayal, Ferrara, Fialkov, Fiore, Ichiki, Illiev, Inoue, Jelic, Jones, Lazio, Maio, Majumdar, Mack, Mesinger, Morales, Parsons, Pen, Santos, Schneider, Semelin, de~Souza, Subrahmanyan, Takeuchi, Vedantham, Wagg, Webster, Wyithe, Datta, and Trott]{Koopmans15_ska}
L.~Koopmans, J.~Pritchard, G.~Mellema, J.~Aguirre, K.~Ahn, R.~Barkana, I.~van Bemmel, G.~Bernardi, A.~Bonaldi, F.~Briggs, A.~G. de~Bruyn, T.~C. Chang, E.~Chapman, X.~Chen, B.~Courty, P.~Dayal, A.~Ferrara, A.~Fialkov, F.~Fiore, K.~Ichiki, I.~T. Illiev, S.~Inoue, V.~Jelic, M.~Jones, J.~Lazio, U.~Maio, S.~Majumdar, K.~J. Mack, A.~Mesinger, M.~F. Morales, A.~Parsons, U.~Pen, M.~Santos, R.~Schneider, B.~Semelin, R.~S. de~Souza, R.~Subrahmanyan, T.~Takeuchi, H.~Vedantham, J.~Wagg, R.~Webster, S.~Wyithe, K.~K. Datta, and C.~Trott.
\newblock {The Cosmic Dawn and Epoch of Reionisation with SKA}.
\newblock \emph{PoS}, AASKA14:\penalty0 001, 2015.
\newblock \doi{10.22323/1.215.0001}.

\bibitem[{Legin} et~al.(2024){Legin}, {Ho}, {Lemos}, {Perreault-Levasseur}, {Ho}, {Hezaveh}, and {Wandelt}]{legin2024_ic}
R.~{Legin}, M.~{Ho}, P.~{Lemos}, L.~{Perreault-Levasseur}, S.~{Ho}, Y.~{Hezaveh}, and B.~{Wandelt}.
\newblock {Posterior sampling of the initial conditions of the universe from non-linear large scale structures using score-based generative models}.
\newblock \emph{\mnras}, 527\penalty0 (1):\penalty0 L173--L178, Jan. 2024.
\newblock \doi{10.1093/mnrasl/slad152}.

\bibitem[Li et~al.(2022)Li, Yang, Chang, Chen, Feng, Xu, Li, and Chen]{li22_superres}
H.~Li, Y.~Yang, M.~Chang, S.~Chen, H.~Feng, Z.~Xu, Q.~Li, and Y.~Chen.
\newblock Srdiff: Single image super-resolution with diffusion probabilistic models.
\newblock \emph{Neurocomputing}, 479:\penalty0 47--59, 2022.
\newblock ISSN 0925-2312.
\newblock \doi{https://doi.org/10.1016/j.neucom.2022.01.029}.
\newblock URL \url{https://www.sciencedirect.com/science/article/pii/S0925231222000522}.

\bibitem[{Magg} et~al.(2022){Magg}, {Reis}, {Fialkov}, {Barkana}, {Klessen}, {Glover}, {Chen}, {Hartwig}, and {Schauer}]{Magg_2022}
M.~{Magg}, I.~{Reis}, A.~{Fialkov}, R.~{Barkana}, R.~S. {Klessen}, S.~C.~O. {Glover}, L.-H. {Chen}, T.~{Hartwig}, and A.~T.~P. {Schauer}.
\newblock {Effect of the cosmological transition to metal-enriched star formation on the hydrogen 21-cm signal}.
\newblock \emph{\mnras}, 514\penalty0 (3):\penalty0 4433--4449, Aug. 2022.
\newblock \doi{10.1093/mnras/stac1664}.

\bibitem[Mesinger(2019)]{mesinger19_review}
A.~Mesinger, editor.
\newblock \emph{The Cosmic 21-cm Revolution}.
\newblock 2514-3433. IOP Publishing, 2019.
\newblock ISBN 978-0-7503-2236-2.
\newblock \doi{10.1088/2514-3433/ab4a73}.
\newblock URL \url{https://dx.doi.org/10.1088/2514-3433/ab4a73}.

\bibitem[{O'Hara} et~al.(2024){O'Hara}, {Dulwich}, {de Lera Acedo}, {Dhandha}, {Gessey-Jones}, {Anstey}, and {Fialkov}]{ohara24_ska}
O.~S.~D. {O'Hara}, F.~{Dulwich}, E.~{de Lera Acedo}, J.~{Dhandha}, T.~{Gessey-Jones}, D.~{Anstey}, and A.~{Fialkov}.
\newblock {Understanding spectral artefacts in SKA-LOW 21-cm cosmology experiments: the impact of cable reflections}.
\newblock \emph{arXiv e-prints}, art. arXiv:2402.04008, Feb. 2024.
\newblock \doi{10.48550/arXiv.2402.04008}.

\bibitem[{Pochinda} et~al.(2024){Pochinda}, {Gessey-Jones}, {Bevins}, {Fialkov}, {Heimersheim}, {Abril-Cabezas}, {de Lera Acedo}, {Singh}, {Sikder}, and {Barkana}]{pochinda24_joint}
S.~{Pochinda}, T.~{Gessey-Jones}, H.~T.~J. {Bevins}, A.~{Fialkov}, S.~{Heimersheim}, I.~{Abril-Cabezas}, E.~{de Lera Acedo}, S.~{Singh}, S.~{Sikder}, and R.~{Barkana}.
\newblock {Constraining the properties of Population III galaxies with multiwavelength observations}.
\newblock \emph{\mnras}, 531\penalty0 (1):\penalty0 1113--1132, June 2024.
\newblock \doi{10.1093/mnras/stae1185}.

\bibitem[Podell et~al.(2024)Podell, English, Lacey, Blattmann, Dockhorn, M{\"u}ller, Penna, and Rombach]{podell24_StableDiffusion}
D.~Podell, Z.~English, K.~Lacey, A.~Blattmann, T.~Dockhorn, J.~M{\"u}ller, J.~Penna, and R.~Rombach.
\newblock {SDXL}: Improving latent diffusion models for high-resolution image synthesis.
\newblock In \emph{The Twelfth International Conference on Learning Representations}, 2024.
\newblock URL \url{https://openreview.net/forum?id=di52zR8xgf}.

\bibitem[{Pritchard} and {Loeb}(2012)]{pritchard12_review}
J.~R. {Pritchard} and A.~{Loeb}.
\newblock {21 cm cosmology in the 21st century}.
\newblock \emph{Reports on Progress in Physics}, 75\penalty0 (8):\penalty0 086901, Aug. 2012.
\newblock \doi{10.1088/0034-4885/75/8/086901}.

\bibitem[{Reis} et~al.(2020){Reis}, {Fialkov}, and {Barkana}]{Reis_2020}
I.~{Reis}, A.~{Fialkov}, and R.~{Barkana}.
\newblock {High-redshift radio galaxies: a potential new source of 21-cm fluctuations}.
\newblock \emph{\mnras}, 499\penalty0 (4):\penalty0 5993--6008, Dec. 2020.
\newblock \doi{10.1093/mnras/staa3091}.

\bibitem[{Reis} et~al.(2021){Reis}, {Fialkov}, and {Barkana}]{Reis_2021}
I.~{Reis}, A.~{Fialkov}, and R.~{Barkana}.
\newblock {The subtlety of Ly {\ensuremath{\alpha}} photons: changing the expected range of the 21-cm signal}.
\newblock \emph{\mnras}, 506\penalty0 (4):\penalty0 5479--5493, Oct. 2021.
\newblock \doi{10.1093/mnras/stab2089}.

\bibitem[{Reis} et~al.(2022){Reis}, {Barkana}, and {Fialkov}]{Reis_2022}
I.~{Reis}, R.~{Barkana}, and A.~{Fialkov}.
\newblock {Shot noise and scatter in the star formation efficiency as a source of 21-cm fluctuations}.
\newblock \emph{\mnras}, 511\penalty0 (4):\penalty0 5265--5273, Apr. 2022.
\newblock \doi{10.1093/mnras/stac411}.

\bibitem[Rombach et~al.(2022)Rombach, Blattmann, Lorenz, Esser, and Ommer]{Rombach22_StableDiffusion}
R.~Rombach, A.~Blattmann, D.~Lorenz, P.~Esser, and B.~Ommer.
\newblock High-resolution image synthesis with latent diffusion models.
\newblock In \emph{Proceedings of the IEEE/CVF Conference on Computer Vision and Pattern Recognition (CVPR)}, pages 10684--10695, June 2022.

\bibitem[{Rouhiainen} et~al.(2024){Rouhiainen}, {M{\"u}nchmeyer}, {Shiu}, {Gira}, and {Lee}]{Rouhiainen24_inpaintSR}
A.~{Rouhiainen}, M.~{M{\"u}nchmeyer}, G.~{Shiu}, M.~{Gira}, and K.~{Lee}.
\newblock {Superresolution emulation of large cosmological fields with a 3D conditional diffusion model}.
\newblock \emph{\prd}, 109\penalty0 (12):\penalty0 123536, June 2024.
\newblock \doi{10.1103/PhysRevD.109.123536}.

\bibitem[Saharia et~al.(2023)Saharia, Ho, Chan, Salimans, Fleet, and Norouzi]{Saharia23_SR3}
C.~Saharia, J.~Ho, W.~Chan, T.~Salimans, D.~J. Fleet, and M.~Norouzi.
\newblock Image super-resolution via iterative refinement.
\newblock \emph{IEEE Transactions on Pattern Analysis and Machine Intelligence}, 45\penalty0 (4):\penalty0 4713--4726, 2023.
\newblock \doi{10.1109/TPAMI.2022.3204461}.

\bibitem[{Scaife}(2020)]{schaife20_ska}
A.~M.~M. {Scaife}.
\newblock {Big telescope, big data: towards exascale with the Square Kilometre Array}.
\newblock \emph{Philosophical Transactions of the Royal Society of London Series A}, 378\penalty0 (2166):\penalty0 20190060, Mar. 2020.
\newblock \doi{10.1098/rsta.2019.0060}.

\bibitem[{Sikder} et~al.(2024){Sikder}, {Barkana}, {Fialkov}, and {Reis}]{Sikder_2023}
S.~{Sikder}, R.~{Barkana}, A.~{Fialkov}, and I.~{Reis}.
\newblock {Strong 21-cm fluctuations and anisotropy due to the line-of-sight effect of radio galaxies at cosmic dawn}.
\newblock \emph{\mnras}, 527\penalty0 (4):\penalty0 10975--10985, Feb. 2024.
\newblock \doi{10.1093/mnras/stad3847}.

\bibitem[{Singh} et~al.(2022){Singh}, {Jishnu}, {Subrahmanyan}, {Udaya Shankar}, {Girish}, {Raghunathan}, {Somashekar}, {Srivani}, and {Sathyanarayana Rao}]{Singh22_saras3}
S.~{Singh}, N.~T. {Jishnu}, R.~{Subrahmanyan}, N.~{Udaya Shankar}, B.~S. {Girish}, A.~{Raghunathan}, R.~{Somashekar}, K.~S. {Srivani}, and M.~{Sathyanarayana Rao}.
\newblock {On the detection of a cosmic dawn signal in the radio background}.
\newblock \emph{Nature Astronomy}, 6:\penalty0 607--617, Feb. 2022.
\newblock \doi{10.1038/s41550-022-01610-5}.

\bibitem[{Smith} et~al.(2022){Smith}, {Geach}, {Jackson}, {Arora}, {Stone}, and {Courteau}]{smith22_galaxydiffusion}
M.~J. {Smith}, J.~E. {Geach}, R.~A. {Jackson}, N.~{Arora}, C.~{Stone}, and S.~{Courteau}.
\newblock {Realistic galaxy image simulation via score-based generative models}.
\newblock \emph{\mnras}, 511\penalty0 (2):\penalty0 1808--1818, Apr. 2022.
\newblock \doi{10.1093/mnras/stac130}.

\bibitem[Sohl-Dickstein et~al.(2015)Sohl-Dickstein, Weiss, Maheswaranathan, and Ganguli]{sohl15_ddpm}
J.~Sohl-Dickstein, E.~Weiss, N.~Maheswaranathan, and S.~Ganguli.
\newblock Deep unsupervised learning using nonequilibrium thermodynamics.
\newblock In F.~Bach and D.~Blei, editors, \emph{Proceedings of the 32nd International Conference on Machine Learning}, volume~37 of \emph{Proceedings of Machine Learning Research}, pages 2256--2265, Lille, France, 07--09 Jul 2015. PMLR.
\newblock URL \url{https://proceedings.mlr.press/v37/sohl-dickstein15.html}.

\bibitem[Song et~al.(2021)Song, Sohl-Dickstein, Kingma, Kumar, Ermon, and Poole]{song2021_sm}
Y.~Song, J.~Sohl-Dickstein, D.~P. Kingma, A.~Kumar, S.~Ermon, and B.~Poole.
\newblock Score-based generative modeling through stochastic differential equations.
\newblock In \emph{International Conference on Learning Representations}, 2021.
\newblock URL \url{https://openreview.net/forum?id=PxTIG12RRHS}.

\bibitem[Vahdat et~al.(2021)Vahdat, Kreis, and Kautz]{vahdat21_sbldm}
A.~Vahdat, K.~Kreis, and J.~Kautz.
\newblock Score-based generative modeling in latent space.
\newblock In M.~Ranzato, A.~Beygelzimer, Y.~Dauphin, P.~Liang, and J.~W. Vaughan, editors, \emph{Advances in Neural Information Processing Systems}, volume~34, pages 11287--11302. Curran Associates, Inc., 2021.
\newblock URL \url{https://proceedings.neurips.cc/paper_files/paper/2021/file/5dca4c6b9e244d24a30b4c45601d9720-Paper.pdf}.

\bibitem[Vaswani et~al.(2017)Vaswani, Shazeer, Parmar, Uszkoreit, Jones, Gomez, Kaiser, and Polosukhin]{vaswani17_transformers}
A.~Vaswani, N.~Shazeer, N.~Parmar, J.~Uszkoreit, L.~Jones, A.~N. Gomez, L.~u. Kaiser, and I.~Polosukhin.
\newblock Attention is all you need.
\newblock In I.~Guyon, U.~V. Luxburg, S.~Bengio, H.~Wallach, R.~Fergus, S.~Vishwanathan, and R.~Garnett, editors, \emph{Advances in Neural Information Processing Systems}, volume~30 of \emph{NIPS 2017}. Curran Associates, Inc., 2017.
\newblock URL \url{https://proceedings.neurips.cc/paper_files/paper/2017/file/3f5ee243547dee91fbd053c1c4a845aa-Paper.pdf}.

\bibitem[{Visbal} et~al.(2012){Visbal}, {Barkana}, {Fialkov}, {Tseliakhovich}, and {Hirata}]{Visbal_2012}
E.~{Visbal}, R.~{Barkana}, A.~{Fialkov}, D.~{Tseliakhovich}, and C.~M. {Hirata}.
\newblock {The signature of the first stars in atomic hydrogen at redshift 20}.
\newblock \emph{\nat}, 487\penalty0 (7405):\penalty0 70--73, July 2012.
\newblock \doi{10.1038/nature11177}.

\bibitem[Wang et~al.(2018)Wang, Girshick, Gupta, and He]{Wang18_selfattention}
X.~Wang, R.~Girshick, A.~Gupta, and K.~He.
\newblock Non-local neural networks.
\newblock In \emph{Proceedings of the IEEE Conference on Computer Vision and Pattern Recognition (CVPR)}, June 2018.

\bibitem[Yang and Mandt(2023)]{yang23_compression}
R.~Yang and S.~Mandt.
\newblock Lossy image compression with conditional diffusion models.
\newblock In A.~Oh, T.~Naumann, A.~Globerson, K.~Saenko, M.~Hardt, and S.~Levine, editors, \emph{Advances in Neural Information Processing Systems}, volume~36 of \emph{NIPS 2023}, pages 64971--64995. Curran Associates, Inc., 2023.
\newblock URL \url{https://proceedings.neurips.cc/paper_files/paper/2023/file/ccf6d8b4a1fe9d9c8192f00c713872ea-Paper-Conference.pdf}.

\bibitem[{Zhao} et~al.(2023){Zhao}, {Ting}, {Diao}, and {Mao}]{Zhao23_21cmdiffusion}
X.~{Zhao}, Y.-S. {Ting}, K.~{Diao}, and Y.~{Mao}.
\newblock {Can diffusion model conditionally generate astrophysical images?}
\newblock \emph{\mnras}, 526\penalty0 (2):\penalty0 1699--1712, Dec. 2023.
\newblock \doi{10.1093/mnras/stad2778}.

\end{thebibliography}

\end{document}